\documentclass[a4paper, amsfonts, amssymb, amsmath, reprint, showkeys, nofootinbib, twoside]{revtex4-1}

\usepackage[english]{babel}
\usepackage[utf8]{inputenc}
\usepackage{gensymb}
\usepackage{amsmath}
\usepackage[colorinlistoftodos, color=green!40, prependcaption]{todonotes}
\usepackage{color}
\usepackage{amsthm}
\usepackage{mathtools}
\usepackage{physics}
\usepackage{xcolor}
\usepackage{graphicx}
\usepackage[left=23mm,right=13mm,top=35mm,columnsep=15pt]{geometry} 

\usepackage{adjustbox}
\usepackage{placeins}
\usepackage[T1]{fontenc}
\usepackage{lipsum}
\usepackage{csquotes}
\usepackage[super]{nth}
\usepackage{textgreek}
\usepackage[utf8]{inputenc}                  
\usepackage{mathtools}
\usepackage{xcolor}
\usepackage{amsmath} 
\usepackage{amsfonts}
\usepackage{amssymb}
\begin{document}
%\title{On-chip spatially-multiplexed degenerate Kerr oscillators \\
%towards photonic Ising machine}

%\title{Photonic spin network towards coherent Ising machine}
\title{Nanophotonic spin-glass for realization of a coherent Ising machine}

\author{Yoshitomo Okawachi$^{1}$, Mengjie Yu$^{1,2}$, Jae K. Jang$^1$, Xingchen Ji$^3$, Yun Zhao$^3$, Bok Young Kim$^1$, \\Michal Lipson$^{1,3}$ \& Alexander~L.~ Gaeta$^{1,3}$}
%    \email[Correspondence email address: ]{a.gaeta@columbia.edu}% Your name
    \affiliation{$^{1}$Department of Applied Physics and Applied Mathematics, Columbia University, New York, NY 10027}
    \affiliation{$^{2}$School of Electrical and Computer Engineering, Cornell University, Ithaca, NY 14853}
    \affiliation{$^{3}$Department of Electrical Engineering, Columbia University, New York, NY 10027}

\date{\today} % Leave empty to omit a date

\begin{abstract}
The need for solving optimization problems is prevalent in a wide range of physical applications, including neuroscience, network design, biological systems, socio-economics, and chemical reactions. Many of these are classified as non-deterministic polynomial-time (NP) hard and thus become intractable to solve as the system scales to a large number of elements. Recent research advances in photonics have sparked interest in using a network of coupled degenerate optical parametric oscillators (DOPO's) to effectively find the ground state of the Ising Hamiltonian, which can be used to solve other combinatorial optimization problems through polynomial-time mapping. Here, using the nanophotonic silicon-nitride platform, we propose a network of on-chip spatial-multiplexed DOPO's for the realization of a photonic coherent Ising machine. We demonstrate the generation and coupling of two microresonator-based DOPO's on a single chip. Through a reconfigurable phase link, we achieve both in-phase and out-of-phase operation, which can be deterministically achieved at a fast regeneration speed of 400 kHz with a large phase tolerance. Our work provides the critical building blocks towards the realization of a chip-scale photonic Ising machine.\end{abstract}

%\keywords{first keyword, second keyword, third keyword}

\maketitle
The processing speed of modern computers are limited by the fact that program memory and data memory share the same bus. While processors have become faster, the overall speed is limited by the data transfer rate, known as the von Neumann bottleneck \cite{Backus78}. With the massive amount of data being produced, the limiting sequential nature of the von Neumann architecture has been exposed. In order to meet the increasing demand for solving certain classes of computation problems that scale exponentially in time and energy, alternative architectures have been explored such as quantum computing and coherent computing \cite{Yamamoto2012,Hamerly}.

\begin{figure}[!b]
\centerline{\includegraphics[width=7cm]{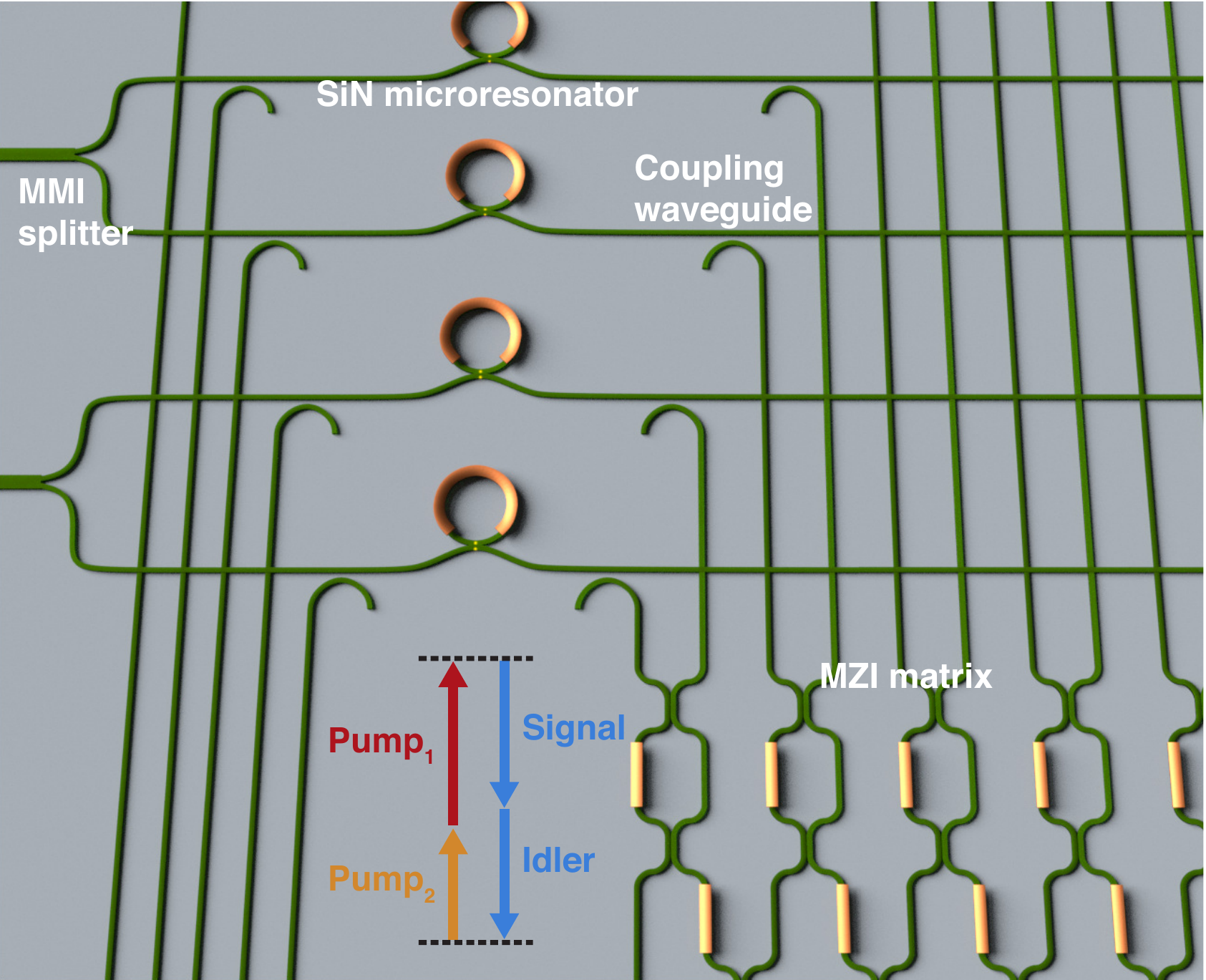}}
\caption{\textbf{Conceptual schematic of a network of coupled DOPO's using SiN microresonators.} Two frequency nondegenerate pumps are injected and the DOPO signal is phase matched and generated via four-wave mixing parametric oscillation. Phase matching conditions result in a bi-phase state for the generated signal. Coupling between DOPO's can be performed using a matrix of reconfigurable Mach-Zehnder interferometers.}
\label{Fig1}
\end{figure}

Recently, there has been considerable interest in using photonic processors to realize a novel form of coherent computing by simulating the Ising model \cite{yamamoto_coherent_2017,InagakiScience,McMahonScience,Guo19}. The Ising model was developed for modeling ferromagnetism and is governed by a Hamiltonian that couples discrete variables that represent spin glasses \cite{Ising,Binder86}. Solving for the ground state of such a system corresponds to solving a non-deterministic polynomial-time (NP) hard computational problem and can provide an architecture for solving other NP-complete problems through polynomial-time mapping to the Ising model \cite{Lucas14}. The Ising Hamiltonian with $N$ spins and no external field is given by $H=-\sum_{ij}^{N} J_{ij}\sigma_i\sigma_j,$ where $J_{ij}$ is the coupling coefficient and $\sigma_i$ corresponds to the projection of the $i^{\text{th}}$ spin along the $z$-axis that can have two states $\pm1$. The physical realization of an Ising machine requires binary degrees of freedom ($i.e.$ spins $\sigma_i$) and reconfigurable coupling ($i.e.$ $J_{ij}$), and initially was studied using a network of injection-locked lasers \cite{Utsunomiya11,Takata12,Takata14,Utsunomiya15,Tradonsky,Babaeian}. More recently, investigations have shown that a network of coupled degenerate optical parametric oscillators (DOPO's) based on the $\chi^{(2)}$ nonlinearity can be used to realize a hybrid temporally multiplexed coherent Ising machine \cite{yamamoto_coherent_2017,Marandi14,InagakiNP,WangPRA,InagakiScience,McMahonScience}, which includes a recent demonstration of a system of 2000 spins \cite{InagakiScience}. The nonlinearity is based on the nonequilibrium phase transition that occurs at the parametric oscillation threshold, resulting in two possible phase states of the DOPO offset by $\pi$, and the couplings between the DOPO's are implemented via measurement-feedback or optical delay lines. By controlling the coupling between these DOPO's, it is possible to achieve more complex, phase-locked output states that encode the ground state of an Ising model. These demonstrations have utilized a time-multiplexed  DOPO system using a 1-km-long fiber ring cavity to simulate the ground state of the Ising model \cite{InagakiScience,McMahonScience}. In addition, extensive experimental and theoretical analysis has been done to characterize the potential performance of such systems \cite{Hamerly,Yamamura17,Haribara17,Bohm18}. Furthermore, alternative approaches towards a coherent Ising machine has been proposed including, opto-electronic oscillators with self-feedback \cite{Bohm19,prabhu19,roques20}, spatial light modulation \cite{Pierangeli}, and dispersive optical bistabilty \cite{Tezak}.

An alternative approach to realize a DOPO is to use a $\chi^{(3)}$ nonlinearity in which a frequency degenerate signal/idler pair is generated via parametric four-wave mixing (FWM) \cite{turitsyn15,Okawachi15,Takesue16,Okawachi16}. Such a scheme has been implemented using silicon nitride (SiN, Si$_3$N$_4$) microresonators and bi-phase state generation has been achieved enabling quantum random-number generation in a chip-scale device (Fig. \ref{Fig1}) \cite{Okawachi15,Okawachi16}. The SiN platform is ideally suited for scalability to an all-photonic network of coupled DOPO's since it is CMOS-process compatible, has low losses in the near-infrared, and allows for dispersion engineering which is crucial for efficient phase-matched nonlinear processes \cite{Gaeta}. Unlike the $\chi^{(2)}$ process, the wavelengths of the pump and signal are spectrally close, allowing for phase-matching of the degenerate signal through dispersion engineering of the fundamental waveguide mode. In addition, unlike traditional FWM which requires operation in the anomalous group-velocity dispersion (GVD) regime, the DOPO requires normal GVD for phase matching \cite{Okawachi15}, which is more readily accessible across a wider range of photonic platforms. Furthermore, the microresonator-based DOPO system allows for simultaneous oscillation of all DOPO's, continuous-wave (cw) operation, and does not rely on long cavity lengths that requires phase stabilization to support the multiple trains of femtosecond pulses as in the time-multiplexing scheme, offering faster computational speeds with lower power consumption in a compact footprint. A recent preliminary study has also reported on numerical simulation of coupled Lugiato-Lefever equations to tackle the Max-Cut problem, further demonstrating the potential capability of SiN platform for optical computing~\cite{Jang18}.

\begin{figure*}[!t]
\centerline{\includegraphics[width=15.5cm]{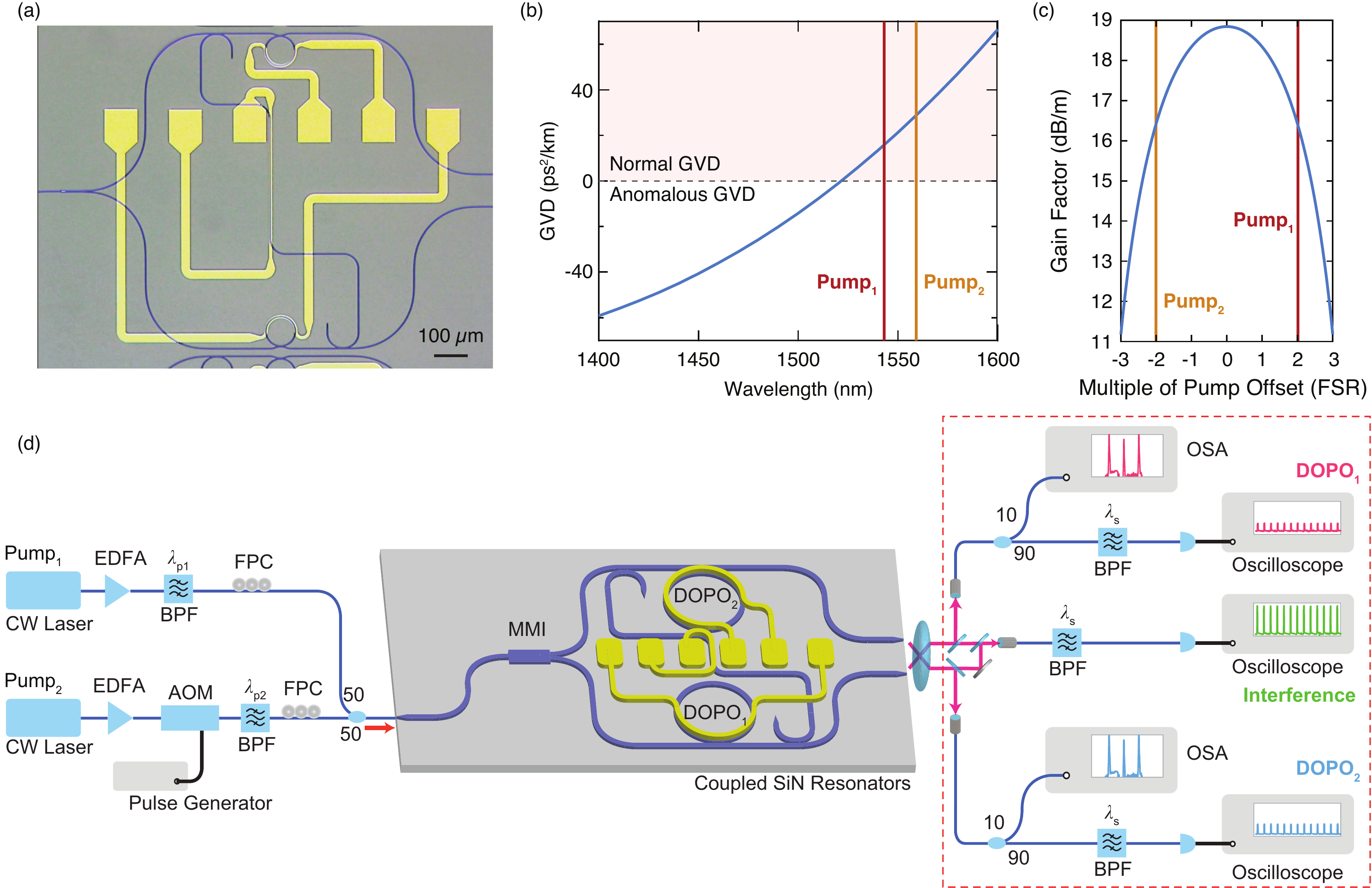}}
\caption{\textbf{Experimental schematic of coupled DOPO's.} (a) Microscope image of the device. The pump waves are split on-chip and sent to DOPO$_1$ (bottom) and DOPO$_2$ (top). A coupling waveguide after DOPO$_1$ is used to send a fraction of the DOPO$_1$ field to DOPO$_2$. (b) Simulated group velocity dispersion (GVD) of the SiN microresonator for the fundamental TE mode. The waveguide cross section is 730$\times$1050 nm. The region of normal GVD is shaded, and the pump wavelengths are indicated with vertical lines. (c) Calculated parametric gain for 1 W of combined pump power. The pump waves are each located 2 FSR's from the degeneracy point. (d) Experimental setup for measurement of coupled-DOPO system. Two frequency nondegenerate pumps are sent into the SiN chip. The output is collected using an aspheric lens and sent to a free-space interferometer to measure the interference signal. EDFA: erbium-doped fiber amplifier, BPF: bandpass filter; AOM: acousto-optic modulator, FPC: fiber polarization controller, MMI: multimode interference splitter, OSA: optical spectrum analyzer.}
\label{Fig2}
\end{figure*}

\begin{figure}[t!]
\centerline{\includegraphics[width=7.5cm]{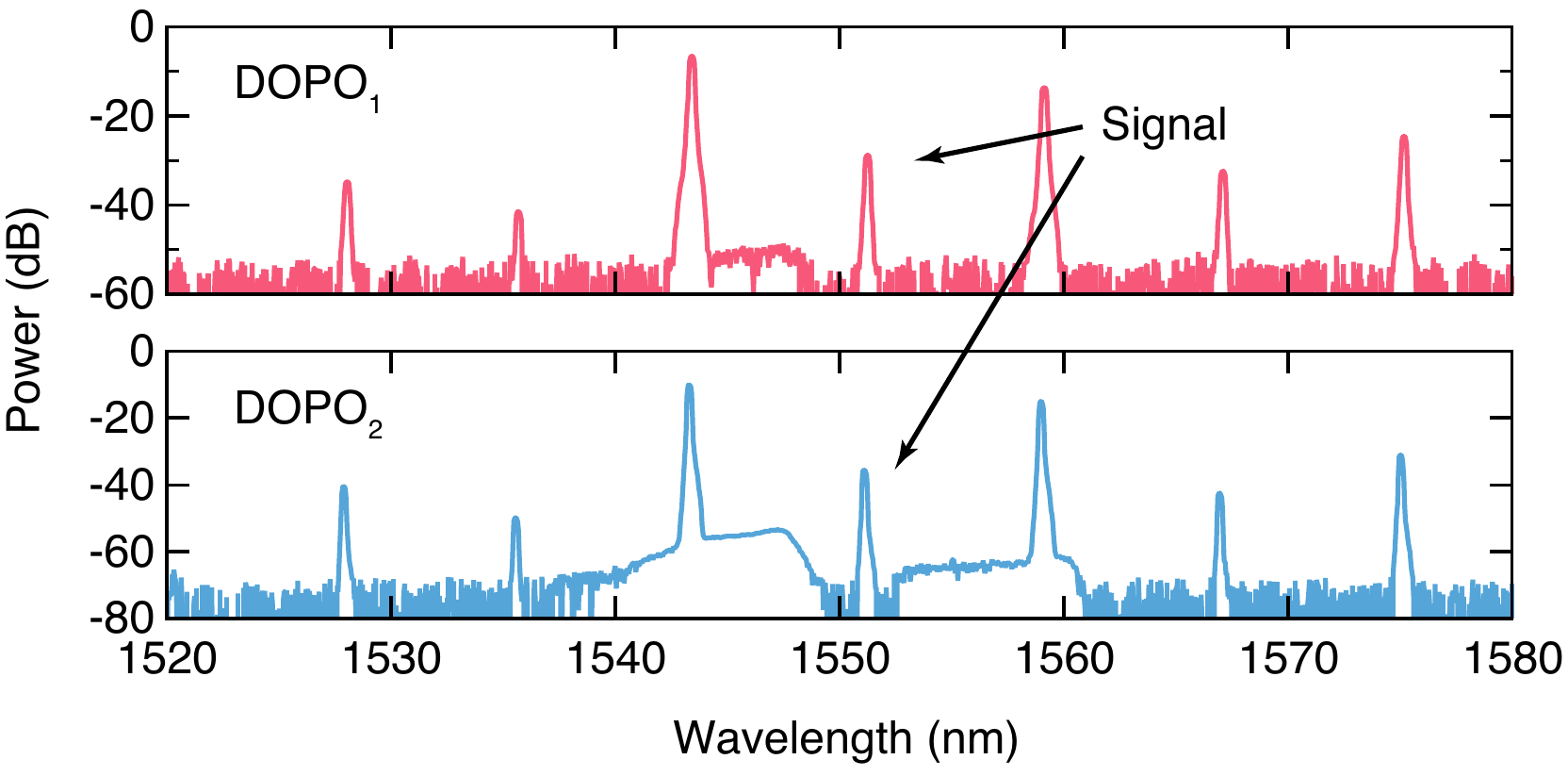}}
\caption{\textbf{Optical spectra of two DOPO signals.} The measured spectra of the primary DOPO (top, DOPO$_1$) and the secondary DOPO (bottom, DOPO$_2$).}
\label{Fig3}
\end{figure}

In this paper, we demonstrate an integrated photonic circuit which consists of spatial-multiplexed DOPO's on a single SiN chip and present the first demonstration of a coupled-DOPO system using continuous-wave pumping. We experimentally show reconfigurability of the coupling phase between the DOPO's through thermal control of the coupling waveguide between the two DOPO's and show interference measurements between the DOPO's indicating in-phase and out-of-phase operation. In our system, we solve at a 400 kHz rate with a convergence time of <310 ns. In addition, we numerically model the coupled-DOPO system and confirm the behavior for in-phase and out-of-phase operation and explore the transition region between the two phase states. A distinct transition region between two states is revealed both numerically and experimentally, suggesting a tremendous phase tolerance of such parametric process. We also describe the scalability of such a system to a large number of oscillators on chip and the challenges to achieving a large-scale integrated system, offering the building blocks toward the realization of a fully photonic coherent Ising machine.

The key components for realizing a network of coupled DOPO's on-chip are, 1) power splitters for routing the pump fields to the microresonators, 2) nonlinear microresonators designed for DOPO generation, and 3) the $N$$\times$$N$ photonic coupling system between the different microresonators. A microscope image of the SiN device is shown in Fig.~\ref{Fig2}(a) and is fabricated using techniques similar to those reported in Luke, \textit{et al.} \cite{Luke13}. For routing the two frequency non-degenerate pumps to the microresonators, we employ an on-chip power splitter using multimode interference (MMI) \cite{Soldano}, where the dimensions of the MMI are designed to allow for 50/50 power splitting ratio. The insertion loss of the MMI splitter is 2 dB. The two SiN microresonators (DOPO$_1$ and DOPO$_2$) have a radius of 45.84 \textmu m, which corresponds to a free spectral range (FSR) of 500 GHz. The loaded quality ($Q$) factors of the microresonators are 630,000. The condition $L_D > L_{NL}$, where $L_D=1/\delta^2|\beta_2|$ is the dispersion length and $L_{NL}=1/2\gamma P$ is the nonlinear length, is critical for achieving maximum gain at the frequency degeneracy point and enabling pure DOPO generation (see Supplementary Information) \cite{Okawachi15}. Here, $\beta_2$ is the GVD parameter, $\gamma$ is the nonlinear parameter, $P$ is the power of each pump, and $\delta$ is the pump frequency offset. Based on simulations using a finite-element mode solver, we use a waveguide cross section of 730$\times$1050 nm such that the two pumps are placed in the normal-GVD regime for the fundamental transverse electric (TE) mode to allow for efficient phase matching and maximum gain at the frequency degeneracy point \cite{Okawachi15}. The simulated GVD is shown in Fig. \ref{Fig2}(b), and the corresponding gain for 1 W of pump power is shown in Fig. \ref{Fig2}(c). The cavity resonance for each microresonator is thermally controlled using integrated platinum resistive microheaters. The coupling between the microresonator and the bus waveguide is designed to have near critical coupling. To compensate for the difference in the resonance frequencies of the two resonators due to the fabrication tolerances in microresonator geometry, we use microheaters above each resonator that allow for electrical control of the resonances via the thermo-optic effect \cite{Cunningham10,Joshi16}. We set the electrical power to the heaters such that the cavity resonances for the two microresonators corresponding to both pump frequencies overlaps. We implement reconfigurable unidirectional coupling between the two DOPO's by using a coupling waveguide that directs a fraction of the DOPO$_1$ output field to the input of DOPO$_2$. The coupling strength of the coupling waveguide is adjusted by designing the separation between the bus waveguide and coupling waveguide. In our device, the ratio between the coupled field from DOPO$_1$ and the DOPO$_2$ field is 0.048 (see Supplementary Information). In order to minimize reflections, we have implemented a taper design on the end facets. We tune the phase coupling between the two DOPO's from in-phase to out-of-phase by thermally tuning the path length using microheaters.

%In order to avoid back-reflections in the coupling waveguide, we implement inverse tapers on both ends \cite{Almeida03}. The separation between the two OPO outputs is chosen to allow for collection using a single aspheric lens.

\begin{figure*}
\centerline{\includegraphics[width=14.5cm]{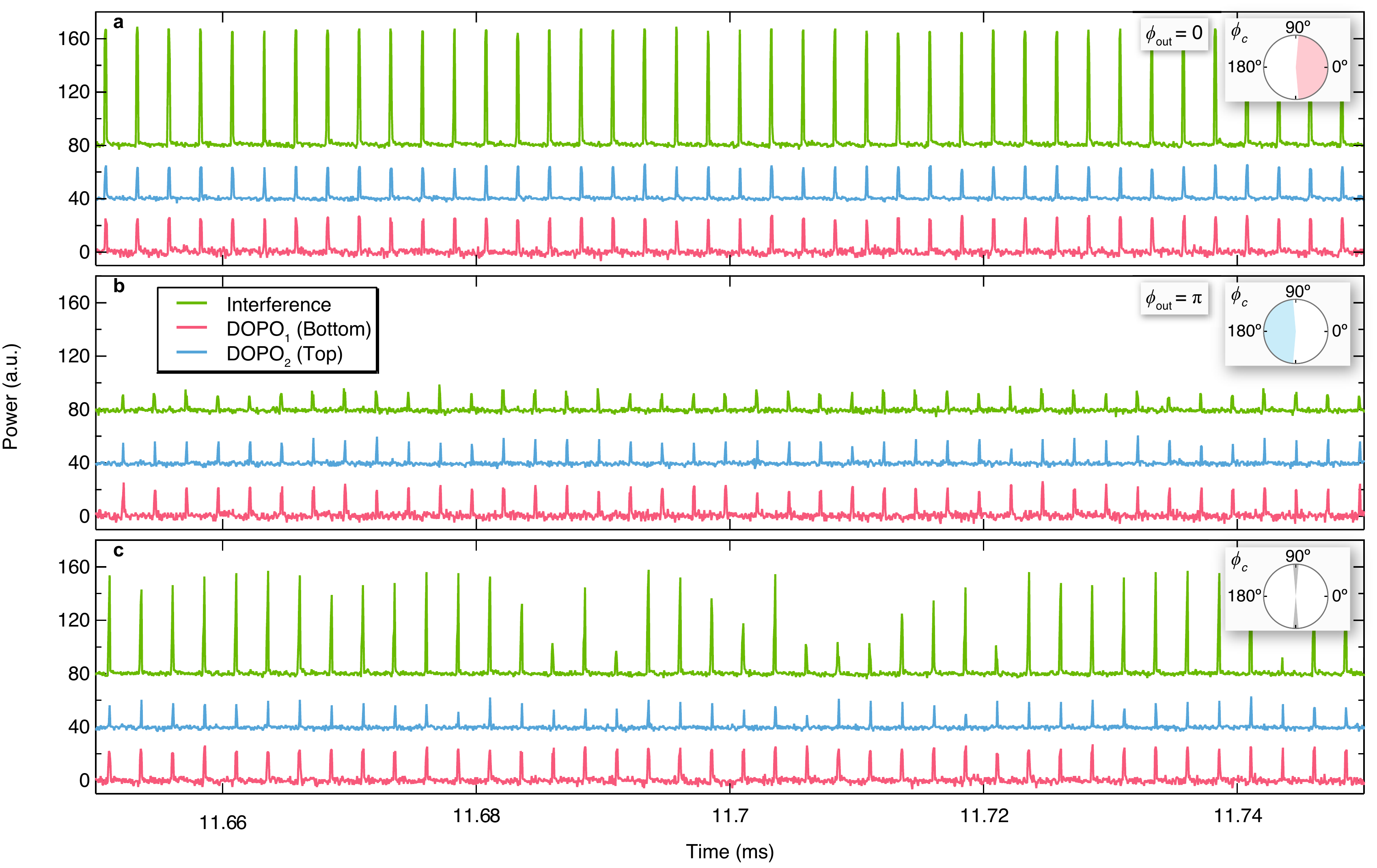}}
\caption{\textbf{Phase characterization of coupled DOPO's.} Temporal ensemble of interference measurements of the combined DOPO's (green) along with measurements of DOPO$_1$ (red) and DOPO$_2$ (blue). Inset shows the phase chart of the coupling phase $\phi_c$. We indicate the coupled DOPO operation regime for in-phase (pink), out-of-phase (blue) and transition regime (grey). (a) Constructive interference (in-phase, $\phi_{\text{out}}=0$) is observed for a heater power of 48.3 mW. (b) Destructive interference (out-of-phase $\phi_{\text{out}}=\pi$) is observed for a heater power of 54.2 mW. (c) Transition from in-phase to out-of-phase operation for heater power of 50.2 mW. For clarity, DOPO$_2$ and the interference are offset from DOPO$_1$ in power by 40 and 80, respectively.}
\label{Fig4}
\end{figure*}

Figure \ref{Fig2}(d) shows the experimental setup for generation and detection of the coupled DOPO signal. The SiN chip is pumped using two frequency nondegenerate cw pump lasers that are offset by $\pm$2 FSR's from the degeneracy point at wavelengths of 1543 nm and 1559.1 nm. In order to ensure that the DOPO builds up from noise for each initiation, we modulate one of the pumps using an acousto-optic modulator with 310-ns pulses at a repetition rate of 400 kHz. For simultaneous degenerate oscillation for both DOPO's, we use 72 mW of combined pump power in each bus waveguide and set the electrical power to the heater in DOPO$_1$ to 13.5 mW such that the resonances of the two microresonators are spectrally overlapped. The two DOPO outputs from the chip are collimated using an aspheric lens and sent to our detection setup [dashed red box in Fig. \ref{Fig2}(a)].

The readout of the coherent phase states of the coupled DOPO system is implemented by directly measuring the interference between the two DOPO's by coupling both outputs into a single fiber collimator and detecting the combined signal on a fast photodiode. The collimated outputs from the DOPO's are combined using a 50/50 beamsplitter and fiber coupled using a collimator. Before combining the beams, a 50/50 beamsplitter is used in each output arm to collect the individual DOPO signals. The signals are each fiber coupled and a 90/10 coupler is used to monitor the time trace and the optical spectrum. The generated DOPO spectra from the microresonators are shown in Fig. \ref{Fig3}. In our interference measurement, we manipulate the coupling phase $\phi_c$ between the DOPO's below threshold by controlling the electrical power sent to the integrated heater while monitoring the time trace. Figure \ref{Fig4} shows the measured temporal interference signal (green) along with DOPO$_1$ (red) and DOPO$_2$ (blue), for three different heater powers. At 48.3 mW, we observe constructive interference between the two DOPO's, corresponding to in-phase operation ($\phi_{\text{out}}=0$) [Fig. \ref{Fig4} (top)]. For 54.2 mW of heater power, we observe destructive interference [Fig. \ref{Fig4} (middle)] corresponding to out-of-phase operation ($\phi_{\text{out}}=\pi$). The transition from in-phase to out-of-phase operation occurs at 50.2 mW of heater power, and we observe both constructive and destructive interference [Fig. \ref{Fig4} (bottom)]. Moreover, we observe in-phase operation ($\phi_{\text{out}}=0$) for heater powers below 48.3 mW and out-of-phase operation ($\phi_{\text{out}}=\pi$) for powers above 54.2 mW, indicating that the DOPO's above threshold operate in-phase or out-of-phase for a continuous range of coupling phases $\phi_c$. The convergence time of the DOPO is well within the pump pulse duration of 310 ns. For comparison, we perform interference measurements on a similar device with no coupling waveguide between DOPO$_1$ and DOPO$_2$ and verify that there is no phase correlation between the two signals (see Supplementary Information).

We theoretically investigate this coupled DOPO system using coupled Lugiato-Lefever equations with two pump waves \cite{Jang18,Lugiato},

%\begin{widetext}
%\begin{align}
%  \!\! \phantom{{t_{R}\frac{\partial A_{-}}{\partial t}}}
%  &\begin{aligned}
%    \mathllap{{t_{R}\frac{\partial E{_1}}{\partial t}}} &= \left(-\frac{\alpha}{2}  - i\delta_{0,1} - iL\frac{\beta_2}{2} \frac{\partial^2}{\partial \tau^2} +i\gamma L|E_1|^2\right)E{_1}(t,\tau) +  \sqrt{\theta} E_{\text{in}}\left(e^{-i\Omega \tau}+e^{i\Omega \tau}\right),
%  \end{aligned}\\
%  &\begin{aligned}
%    \mathllap{{t_{R}\frac{\partial E{_2}}{\partial t}}} &= \left(-\frac{\alpha}{2}  - i\delta_{0,2} - iL\frac{\beta_2}{2} \frac{\partial^2}{\partial \tau^2} +i\gamma L|E_2|^2\right)E{_2}(t,\tau) +  \sqrt{\theta} E_{\text{in}}\left(e^{-i\Omega \tau}+e^{i\Omega \tau}\right)+\kappa E_1(t,\tau),
%  \end{aligned}
%\end{align}
%\smallskip
%\end{widetext}

%\small
%\begin{align}
%  \!\! \phantom{{t_{R}\frac{\partial E_{1}}{\partial t}}}
%  &\begin{aligned}
%    \mathllap{{t_{R}\frac{\partial E{_1}}{\partial t}}} &= \left(-\frac{\alpha}{2}  - i\delta_{0,1} - iL\sum_{n \ge 2} \frac{\beta_n}{n!} \frac{\partial^2}{\partial \tau^2} +i\gamma L|E_1|^2\right)E{_1}(t,\tau) \\
%      &\quad     +  \sqrt{\theta} E_{\text{in}}\left(e^{-i\Omega \tau}+e^{i\Omega \tau}\right),
%  \end{aligned}\\
%  &\begin{aligned}
%    \mathllap{{t_{R}\frac{\partial E{_2}}{\partial t}}} &= \left(-\frac{\alpha}{2}  - i\delta_{0,2} - iL\sum_{n \ge 2} \frac{\beta_n}{n!} \frac{\partial^2}{\partial \tau^2} +i\gamma L|E_2|^2\right)E{_2}(t,\tau) \\
%      &\quad     +  \sqrt{\theta} E_{\text{in}}\left(e^{-i\Omega \tau}+e^{i\Omega \tau}\right)+\kappa E_1(t,\tau),
%  \end{aligned}
%\end{align}
%\normalsize

\begin{widetext}
\begin{equation*}
\begin{aligned}[c]
t_{\mathrm{R}}\frac{\partial E_1}{\partial t} = &\bigg[-\alpha-i\delta_0- \delta_1\frac{\partial}{\partial\tau} +iL\sum_{k\geq 2}\frac{\beta_k}{k!}\bigg(i\frac{\partial}{\partial\tau}\bigg)^k +i\gamma L|E_1(t,\tau)|^2  \bigg]E_1(t,\tau) + \sqrt{\theta}A_{\mathrm{in}}\big(e^{-i\Omega_0\tau} + e^{i\Omega_0\tau} \big)\\
t_{\mathrm{R}}\frac{\partial E_2}{\partial t} = &\bigg[-\alpha-i\delta_0- \delta_1\frac{\partial}{\partial\tau} +iL\sum_{k\geq 2}\frac{\beta_k}{k!}\bigg(i\frac{\partial}{\partial\tau}\bigg)^k+i\gamma L|E_2(t,\tau)|^2 \bigg]E_2(t,\tau) + \sqrt{\theta}A_{\mathrm{in}}\big(e^{-i\Omega_0\tau} + e^{i\Omega_0\tau}\big)\\
&+\kappa E_1(t,\tau),
\end{aligned}
\end{equation*}
\end{widetext}

%\small
%\begin{align}
%  \!\! \phantom{{t_{R}\frac{\partial E_{1}}{\partial t}}}
%  &\begin{aligned}
%    \mathllap{{t_{R}\frac{\partial E{_1}}{\partial t}}} &= \left(-\frac{\alpha}{2}  - i\delta_{0,1} - iL\frac{\beta_2}{2} \frac{\partial^2}{\partial \tau^2} +i\gamma L|E_1|^2\right)E{_1}(t,\tau) \\
%      &\quad     +  \sqrt{\theta} E_{\text{in}}\left(e^{-i\Omega \tau}+e^{i\Omega \tau}\right),
%  \end{aligned}\\
%  &\begin{aligned}
%    \mathllap{{t_{R}\frac{\partial E{_2}}{\partial t}}} &= \left(-\frac{\alpha}{2}  - i\delta_{0,2} - iL\frac{\beta_2}{2} \frac{\partial^2}{\partial \tau^2} +i\gamma L|E_2|^2\right)E{_2}(t,\tau) \\
%      &\quad     +  \sqrt{\theta} E_{\text{in}}\left(e^{-i\Omega \tau}+e^{i\Omega \tau}\right)+\kappa E_1(t,\tau),
%  \end{aligned}
%\end{align}
%\normalsize
\begin{figure}[!b]
\centerline{\includegraphics[width=6.5cm]{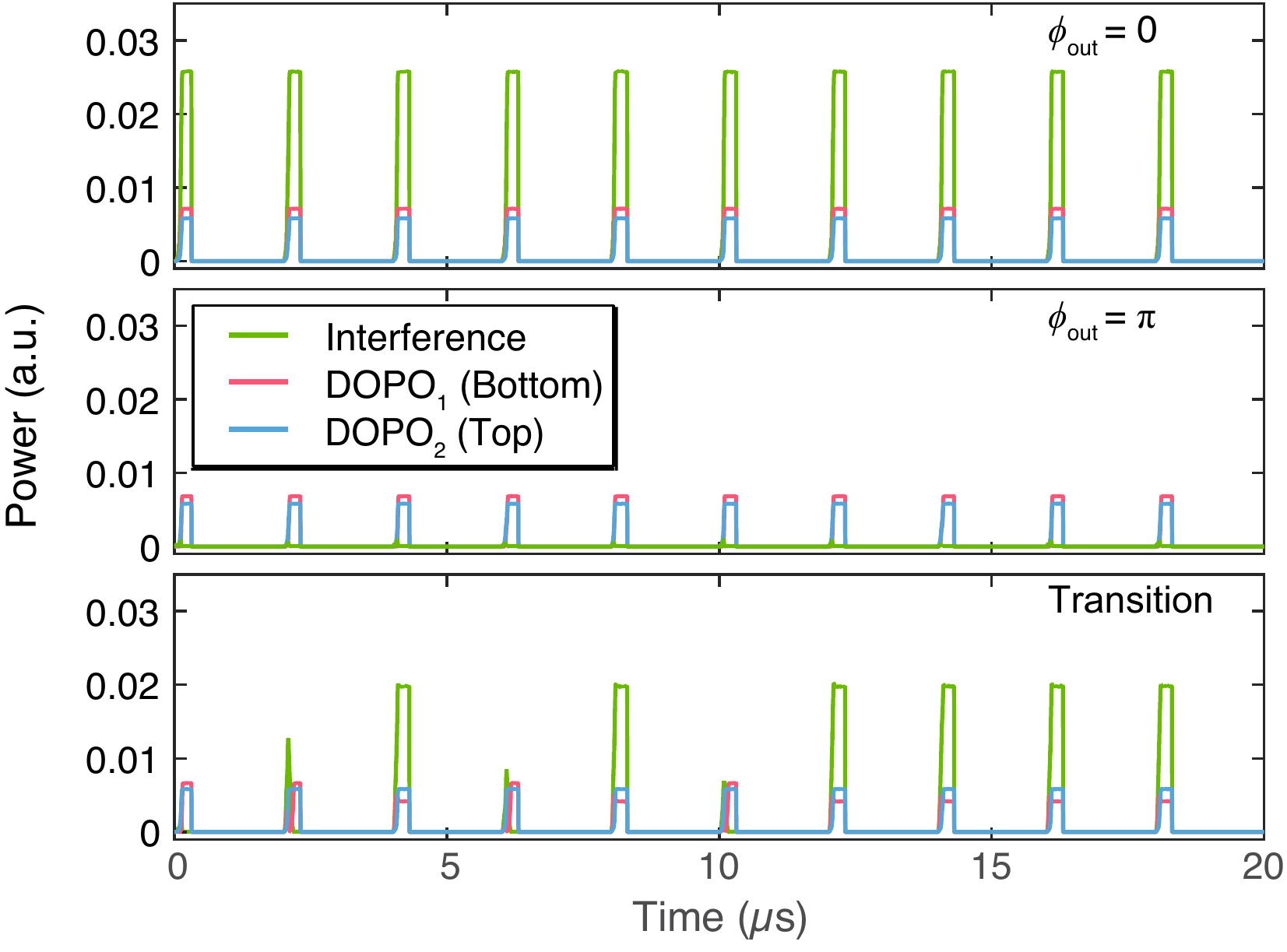}}
\caption{\textbf{Numerical modeling of coupled DOPO system.} Simulated temporal interference between two DOPO's with unidirectional coupling. The plot shows the combined DOPO (green) and the individual DOPO's, DOPO$_1$ (red) and DOPO$_2$ (blue) for coupling phase $\phi_c$ of $0\degree$ (in-phase, $\phi_{\text{out}}=0$), $180\degree$ (out-of-phase, $\phi_{\text{out}}=\pi$), and $90\degree$ (transition from in-phase to out-of-phase) from top to bottom.}
\label{Fig5}
\end{figure}

\noindent where $t_R$ is the roundtrip time in the resonator, $\alpha$ is the total roundtrip loss, $\delta_{0}$ is the effective phase detuning, $\delta_{1}$ is the mode-dependent detuning, $\Omega_0$  corresponds to the pump offset frequency from the degeneracy point, $\theta$ is the transmission coefficient between the resonator and the bus waveguide, $L$ is the cavity length, $\gamma$ is the nonlinear parameter, and $\beta_k$ corresponds to the $k$\textsuperscript{th}-order dispersion coefficients of the Taylor expansion of the propagation constant. Here, $\tau$ represents the temporal coordinate within the time scale of a single round trip and $t$ represents the long-time-scale evolution over many round trips.  The term with $A_{\text{in}}$ describes the bichromatic pump waves, and $\kappa$ represents the complex coupling coefficient from the primary DOPO (DOPO$_1$) to the secondary DOPO (DOPO$_2$) (see Supplementary Information).  We add a noise of one photon per spectral mode with random phase onto the pump \cite{Dudley02}. The simulation parameters are similar to that of our experiment, including the coupling between the DOPO's being unidirectional. Our simulations (Fig. \ref{Fig5}) show oscillation behavior for three different values of the coupling phase $\phi_c$. For $\phi_c=0\degree$, constructive interference between the DOPO's is favored [Fig. \ref{Fig5} (top)], indicating that the DOPO's oscillate in-phase. We observe similar in-phase behavior for $-70\degree<\phi_c<70\degree$. In contrast, for $\phi_c=180\degree$, we observe destructive interference [Fig. \ref{Fig5} (middle)] from the DOPO's oscillating $\pi$ out-of-phase. Likewise, similar out-of-phase behavior is observed for $110\degree<\phi_c<250\degree$. These predictions are consistent with our experimental results. In addition, we have numerically explored the transition region between in-phase and out-of-phase operation. Figure \ref{Fig5} (bottom) shows the simulated interference for $\phi_c=90\degree$. Here, we observe that the secondary DOPO no longer oscillates out-of-phase with respect to the primary DOPO and becomes frustrated with the oscillation becoming uncorrelated, which is also consistent with our experimental observations for the transition region (heater power of 50.2 mW).  Work is ongoing to determine the degree of phase tolerance as the number of DOPO's is increased.

%\begin{figure}[!t]
%\centerline{\includegraphics[width=8cm]{Fig5.eps}}
%\caption{\textbf{Numerical modeling of coupled OPO system.} Simulated temporal interference between two OPO's with unidirectional coupling. The plot shows the combined OPO (green) and the individual OPO's, OPO$_1$ (red) and OPO$_2$ (blue) for operation in-phase ($\phi_{\text{out}}=0$), out-of-phase ($\phi_{\text{out}}=\pi$), and the transition from in-phase to out-of-phase from top to bottom, respectively.}
%\label{Fig5}
%\end{figure}

%Lastly, we have developed a preliminary 4-OPO system and have observed degenerate OPO generation in the microresonator. We preform measurements on a 4-OPO system as shown in Fig. \ref{Fig6}(a). MMI splitters in series are used to route the pumps to four SiN microresonators. Unidirectional coupling is implemented between each adjacent OPO by using a coupling waveguide between the bus waveguides. To maintain a single plane, the coupling waveguide between OPO$_4$ and OPO$_1$ crosses the bus waveguide for OPO$_2$, OPO$_3$, and the coupling waveguide between OPO$_1$ and OPO$_2$. Figure \ref{Fig6}(b) shows the generated spectrum for one of the OPO's.

\begin{figure}[!t]
 \centerline{\includegraphics[width=8cm]{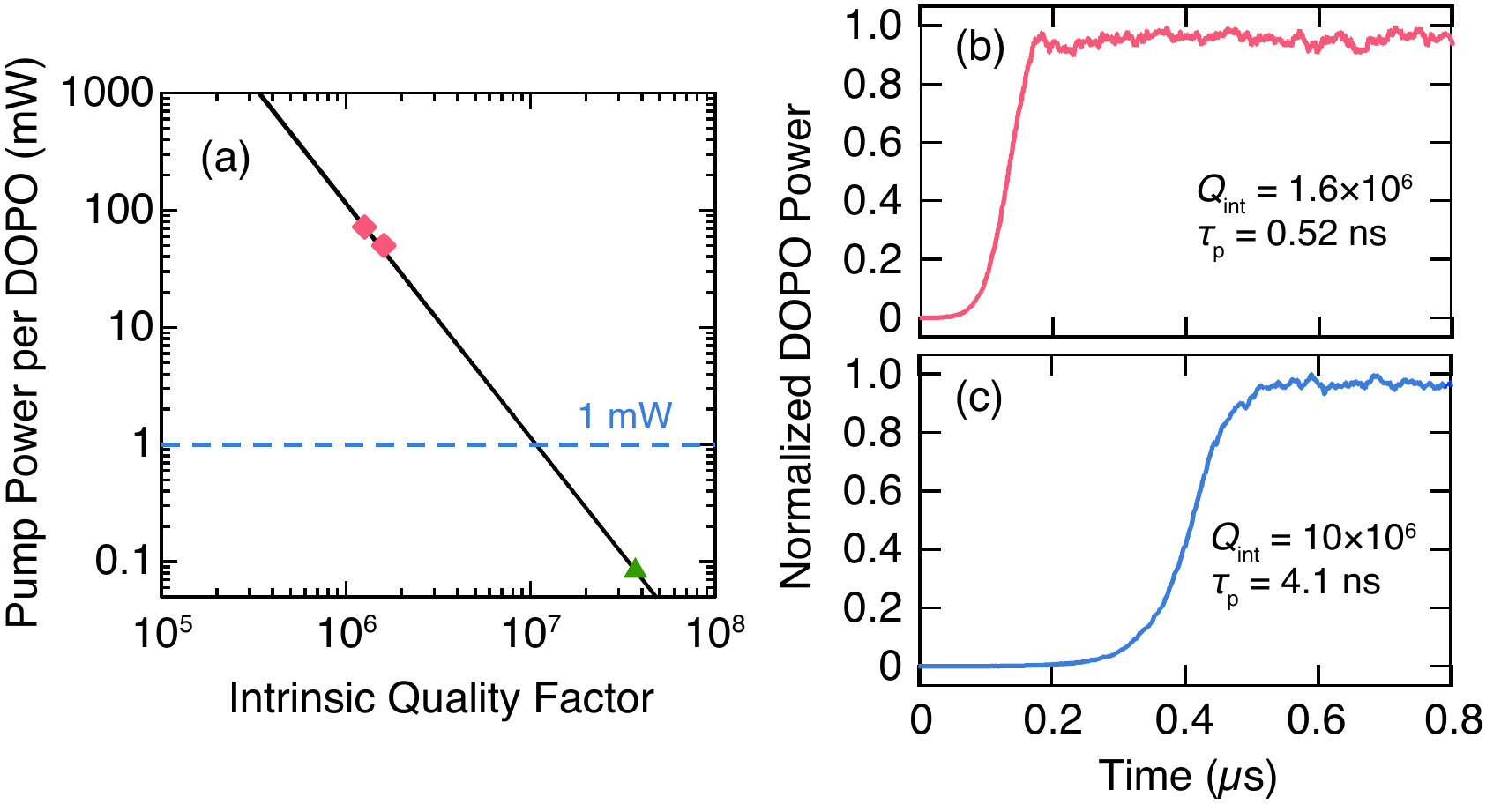}}
\caption{\textbf{Scaling to larger number of coupled-DOPO's.} (a) Combined pump power required for single DOPO as a function of the intrinsic $Q$-factor of the microresonator. The two different pump powers (and the corresponding $Q$) used in our experiments is denoted with diamonds. Triangle (green) denotes pump power based on $Q$ of state-of-the-art SiN microresonator \cite{Ji17}. Simulated DOPO convergence for (b) $Q_{\text{int}}=1.26\times10^6$  which corresponds to a cavity lifetime $\tau_{\text{p}}=520$ ps, and (c) $Q_{\text{int}}=10\times10^6$ which corresponds to $\tau_{\text{p}}=4.1$ ns, for critical coupling.}
    \label{Fig6}
  \end{figure}

Lastly, we discuss the potential of a large-scale DOPO photonic system in light of power consumption and computing speed. Figure \ref{Fig6}(a) shows the combined pump power required for a single microresonator-based DOPO to oscillate as a function of the intrinsic $Q$-factor of the SiN microresonator for critical coupling. The measured values for the current device and a similar single DOPO device are denoted with a diamond (see Supplementary Information). For a $Q$-factor of 10 million, oscillation can occur with 1 mW of combined pump power, implying that 1000 DOPO's can be pumped simultaneously with an on-chip optical power of 1 W, offering promise for scaling to a large number of DOPO's. Here, since the power scales as 1/$Q^2$ and the lifetime scales as $Q$, the required energy for the entire computation scales as 1/$Q$. With reduction of surface roughness, $Q$-factors of 37 million have been achieved in high-confinement SiN microresonators \cite{Ji17}, which could further reduce the pump power to 80 {\textmu}W per DOPO [labeled with a triangle in \ref{Fig6}(a)], at the expense of computing rate. In addition, Figs. \ref{Fig6}(b) and (c) shows the simulated DOPO oscillation time for two different $Q$-factors at near-threshold pump powers. For a single DOPO in a microresonator with an intrinsic-$Q$ of 1.26 million (our experiment), we observe a convergence time of 174 ns with a cavity lifetime of 520 ps based on our numerical model. For a microresonator with an intrinsic-$Q$ of 10 million, we observe convergence times of 540 ns. Experimentally in our 2-DOPO system, we achieve a convergence time <310 ns with a computing rate of 400 kHz . As a comparison, the anneal time of a time-multiplexed 4-DOPO system shown by Marandi \textit{et al.} is 180 {\textmu}s with a computing rate of 1 kHz \cite{Marandi14}. This suggests that the integrated photonics platform offers promise for accelerating the computing time of an Ising model. Furthermore, as the number of DOPO's in the network increases, the anneal time which corresponds to the pump turn-on time must be controlled to slow down the DOPO dynamics to prevent freeze-out effects that prevent the system from reaching the ground-state solution \cite{Bohm18,Hamerly}. More specifically for our system of microresonator-based DOPO's, it has been reported elsewhere that the rate at which the DOPO's are tuned into resonance is a decisive parameter that determines the success probability of finding the ground-state solution (see Supplementary Information)~\cite{Jang18}. As the pump-to-resonance detuning dictates the power build-up within the microresonator, this observation is consistent with previous studies where the pump power was ramped up at a controlled rate to improve the performance of their systems~\cite{InagakiScience}. In other words, the detuning is the more natural control parameter for our microresonator-based DOPO's which must be carefully tuned. This fact will be subject to a more comprehensive investigation and reported elsewhere. We also observe that the oscillation threshold is reduced for a 2-DOPO system for both in-phase and out-of-phase couplings~\cite{Jang18} which leads to apparent faster convergence times as compared to a single DOPO. Investigations are ongoing for the optimal turn-on time as the problem size further increases for our spatial-multiplexed DOPO system. Finally, we have developed a preliminary 4-DOPO system and have observed DOPO generation in a single microresonator as shown in the Supplementary information. Future work will implement reconfigurable phase coupling between the adjacent DOPO's via integrated heaters.

In conclusion, we demonstrate phase reconfigurable all-optical coupling between microresonator-based DOPO's in an integrated silicon nitride platform. We observe that the system is highly tolerant to the coupling phase between the DOPO's, offering flexibility in setting up the system when scaling to larger number of DOPO's. Since the system does not rely on time-multiplexing, and the computation time is comparable to the build-up time of a single DOPO, it is possible to rapidly test the fidelity of the final state. To enable reconfigurable coupling between arbitrary DOPO's, a fraction of the DOPO power will be sent to a matrix of $2N$$\times$$2N$ Mach-Zehnder interferometers (MZI's) to perform $N$$\times$$N$ arbitrary linear transformations \cite{Reck94,Clements16,Shen17}. To achieve this on a single device layer, intersections between waveguides are required. Our transmission measurements (see Supplementary Information) indicate 0.45 dB loss per intersection which can be further optimized. In addition, we are investigating a multi-layer design \cite{MacFarlane} where the coupling matrix based on MZI's is located in a different material plane which uses silicon for the waveguides. Due to the larger thermo-optic effect in silicon, the MZI's can be further miniaturized such that each MZI element resides on a footprint of 100 \textmu m $\times$100 \textmu m \cite{Watts13}. Such a design would enable all-to-all coupling among the DOPO's with arbitrary weightings and the resulting system should in principle allow the implementation of an arbitrary Hamiltonian. Our results provide the initial building blocks toward the realization of a photonic coherent Ising machine for solving combinatorial optimization problems.

\section*{Methods}
%\noindent \textbf{Device information}

\noindent \textbf{Experimental setup} The setup is shown in Fig. \ref{Fig2}(a). Two tunable cw lasers (New Focus Velocity) are amplified using an erbium-doped fiber amplifier (EDFA) to use as pumps. The high-wavelength pump is modulated using an acousto-optic modulator (AOM), where the modulation depth and frequency are carefully chosen such that the DOPO signal reaches the noise level each time the AOM turned off. In order to suppress the amplified spontaneous emission from the EDFA at the DOPO wavelength, we use 9-nm wide bandpass filters after each EDFA. Fiber polarization controllers are used in each arm to set the input polarization to TE. For the chip output, the individual DOPO arms and the interferometer output is collected using a fiber collimator and the DOPO wavelength is filtered using a tunable bandpass filter with a 0.8 nm bandwidth and sent to an InGaAs amplifier photodetector (150-MHz bandwidth) and sent to a 1-GHz real-time oscilloscope.\\
    \\
\noindent \textbf{Acknowledgements} This work was supported by Army Research Office (ARO) (grant W911NF-17-1-0016), National Science Foundation (NSF) (grant CCF-1640108), Semiconductor Research Corporation (SRC) (grant SRS 2016-EP-2693-A), and Air Force Office of Scientific Research (AFOSR) (grant FA9550-15-1-0303). This work was performed in part at the Cornell Nano-Scale Facility, which is a member of the National Nanotechnology Infrastructure Network, supported by the NSF, and at the CUNY Advanced Science Research Center NanoFabrication Facility. We also acknowledge useful discussions with A. Farsi, C. Joshi, C. Joshi, A. Mohanty, S. Ramelow, and L. Shao.
\\
\\
\noindent \textbf{Author contributions}
Y.O. and M.Y. designed the devices and experiment. Y.O., M.Y. and Y.Z. performed the experiments. J.K.J. performed numerical modeling. X.J. fabricated the devices. Y.O., M.Y., Y.Z., and B.Y.K. performed characterization of the DOPO devices. Y.O. and M.Y. analyzed the data. Y.O. prepared the manuscript in discussion with all authors. M.L. and A.L.G. supervised the project.
Y.O. and M.Y. contributed equally to this work.
\\
\\
\noindent \textbf{Competing interests}
The authors declare that they have no
competing financial interests.
\\

\noindent \textbf{Code availability}
The modeling is described in the supplementary information and the code is available from the corresponding author upon reasonable request.
\\

\noindent \textbf{Data availability}
The data that support the plots within this paper and other findings of this study are available from the corresponding author upon reasonable request.
\\
\noindent \textbf{Correspondence}
Correspondence and requests for materials
should be addressed to A.L.G.~(email: a.gaeta@columbia.edu).

\bibliography{DOPO}

\end{document}